\documentclass[showkeys,preprintnumbers,floatfix,amsmath,a4paper,nofootinbib,amssymb]{revtex4}
    \usepackage[margin=0.7in]{geometry}
    \usepackage[parfill]{parskip}
    \usepackage[utf8]{inputenc}
    \usepackage{hyperref}
    \usepackage{graphicx}
    \usepackage[dvipsnames]{xcolor}
    \usepackage{amsmath,amssymb,amsfonts,amsthm}
    \usepackage[textwidth=1cm,textsize=smaller]{todonotes}
  
    \newcommand{\beginsupplement}{%
        \setcounter{table}{0}
        \renewcommand{\thetable}{S\arabic{table}}%
        \setcounter{figure}{0}
        \renewcommand{\thefigure}{S\arabic{figure}}%
     }

\begin{document}
\title{Quota-based debiasing can decrease representation\\of already underrepresented groups}
\author{Ivan Smirnov}
\email{ibsmirnov@hse.ru}
\affiliation{National Research University Higher School of Economics}
\author{Florian Lemmerich}
\email{florian.lemmerich@cssh.rwth-aachen.de}
\affiliation{RWTH Aachen University}
\author{Markus Strohmaier}
\email{markus.strohmaier@cssh.rwth-aachen.de}
\affiliation{RWTH Aachen University and GESIS - Leibniz Institute for the Social Sciences}
\date{June 12th 2020}

\begin{abstract}
Many important decisions in societies such as school admissions, hiring, or elections are based on the selection of top-ranking individuals from a larger pool of candidates. This process is often subject to biases, which typically manifest as an under-representation of certain groups among the selected or accepted individuals. 
The most common approach to this issue is debiasing, for example via the introduction of quotas that ensure proportional representation of groups with respect to a certain, often binary attribute. 
Cases include quotas for women on corporate boards or ethnic quotas in elections. 
This, however, has the potential to induce changes in representation with respect to other attributes. For the case of two correlated binary attributes we show that quota-based debiasing based on a single attribute can worsen the representation of already underrepresented groups and decrease overall fairness of selection. We use several data sets from a broad range of domains from recidivism risk assessments to scientific citations to assess this effect in real-world settings. Our results demonstrate the importance of including \emph{all} relevant attributes in debiasing procedures and that more efforts need to be put into eliminating the root causes of inequalities as purely numerical solutions such as quota-based debiasing might lead to unintended consequences.
\end{abstract}
\keywords{quota, debiasing, algorithmic fairness} 
\maketitle

Selection of top-ranked individuals from a larger pool of candidates is a ubiquitous mechanism for decision making. In many countries, school admission is determined by the selection of top graduates based on their test scores. Elections are, in general, a selection of top candidates based on the number of votes they get. Hirings and promotions are essentially processes of choosing top individuals from a limited pool of candidates based on an implicit ranking of their skills. 
Such processes are known to be affected by biases. For example, hiring decisions have been found to be biased with respect to gender \cite{moss2012science,williams2015national} and ethnicity \cite{quillian2017meta,bessudnov2020ethnic}. Given the crucial role that selection processes play in shaping our everyday life and their potentially high-stakes consequences, eliminating -- or at least limiting -- undesirable biases is extremely important. 

The most common solution is to introduce quotas that ensure proportional representation of groups with respect to a certain, often binary attribute. Examples include, among many others, quotas for women on corporate boards \cite{terjesen2016board}, ethnic quotas in elections \cite{krook2014electoral}, and quotas based on states of origin in university admissions \cite{okoroma2008admission}. 
While successfully eliminating underrepresentation with respect to one attribute, quotas typically ignore changes in the representation with respect to other attributes. This can lead to unintended consequences and can even decrease the representation of already underrepresented groups.

\section*{The Debiasing Paradox}

We define the \emph{Debiasing Paradox} to describe paradoxical situations, in which interventions that reduce bias for groups defined by a property can further decrease the representation of an already underrepresented subgroup.
This paradox occurs when other potentially sensitive -- but invisible or ignored -- attributes are correlated with the attribute that is used for debiasing, which can happen quite naturally in real world settings. One example is the pay gap between women and men that could partially be explained by the wage penalty for mothers \cite{kleven2019children,budig2001wage}. In this case, two attributes -- “being a woman” and “taking care of children” -- are correlated and both could have negative effects on salary. Debiasing on the first attribute might lead to unintended side effects for some minority groups, i.e. women who are not taking care of children or men who do. Next, we explore the Debiasing Paradox both theoretically and empirically.

\begin{figure}[t]
\centering
\includegraphics[width=\linewidth]{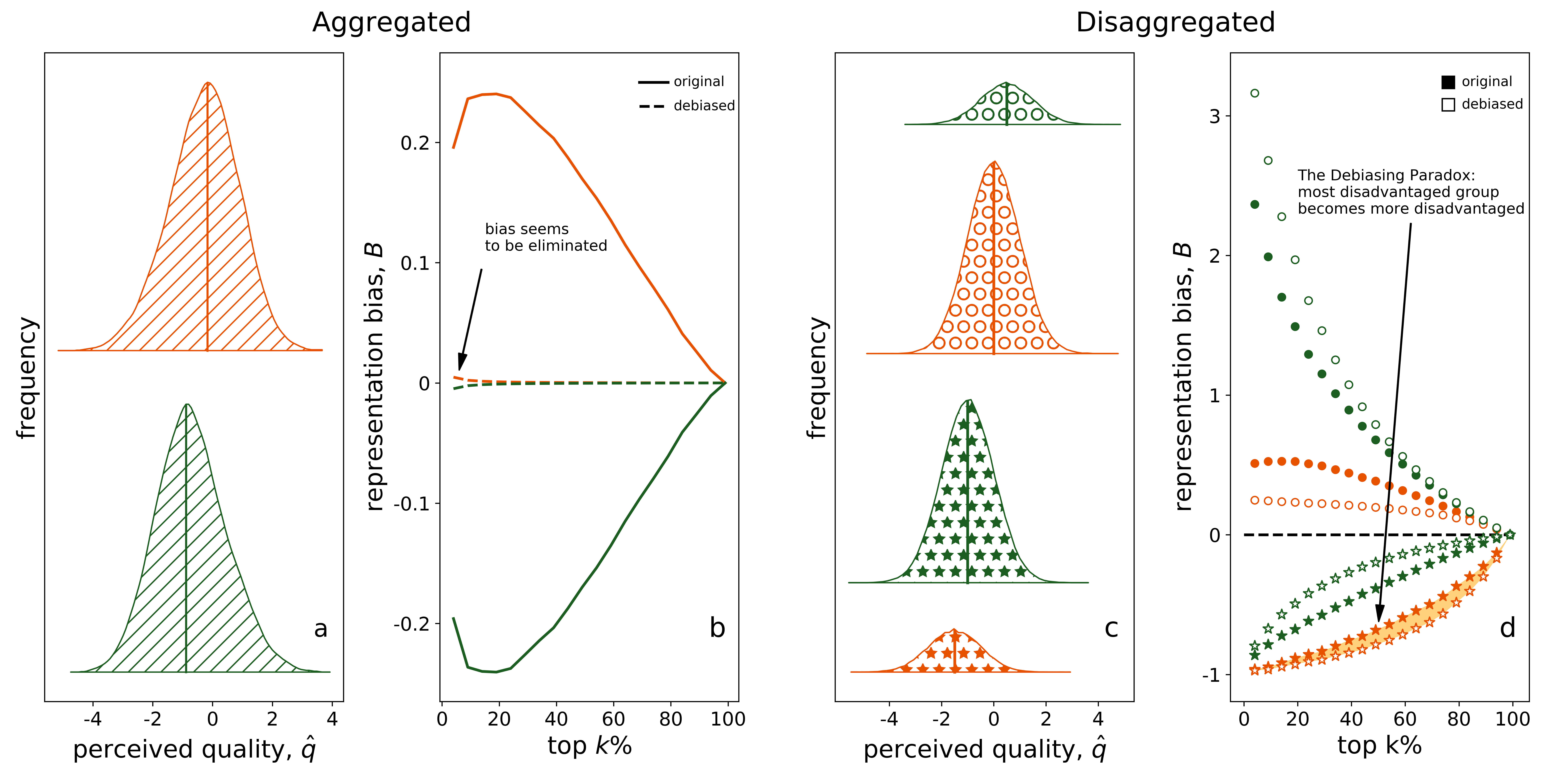}
\caption{
\textbf{The Debiasing Paradox}. 
If only one attribute (\textit{color}) is considered then orange entities appear to have an advantage as their average perceived quality is higher (a). In fact, being orange is a disadvantage by construction (c). In this case, while debiasing on \textit{color} seems to eliminate color bias (b), it, in fact, affects various subgroups differently (d). In particular, it worsens the representation of the already most disadvantaged group of orange stars. We call this effect the debiasing paradox.
}
\label{fig:image}
\end{figure}

\section*{Theoretical Model}
We present a theoretical model with correlated binary attributes to demonstrate that debiasing can paradoxically worsen the representation of the most disadvantaged group if a second hidden attribute is taken into account. This can for example happen if a discrepancy between aggregated and disaggregated data is observed, cf. Simpson's paradox \cite{blyth1972simpson}).

We consider a world populated by social entities (individuals) that have a certain inherent quality $q$ for a task, such that $q \sim N(0,1)$ is normally distributed. The entities have one attribute (e.g., \textit{color:} green or orange) that is visible to a public, and another attribute (e.g., \textit{shape:} stars or circles) that is invisible or ignored. Both attributes (color and shape) are correlated with each other. For simplicity, we assume that there are equal numbers of stars and circles ($N$) as well as equal numbers of green and orange entities. That is, there are $f * N$ green circles and $f * N$ orange stars, where $0 < f < 0.5$.

In this setting, we now consider biases in the perception of this quality. We assume that instead of the real quality $q$ the selection of top candidates is based on perceived quality $\hat{q}$, where $\hat{q}_i = q_i - d_{color} * I_i^{color} - d_{shape} * I_i^{shape}$, i.e., the perceived quality is lower than the real quality for entities of a particular color and a particular shape. Here, the indicator function $I_i^{color}$ is $1$ for green and $0$ for orange entities, $I_i^{shape}$ is $1$ for stars and $0$ for circles, $d_{color}$ and $d_{shape}$ are fixed biases. We then explore how debiasing on the visible attribute \textit{color} affects the representation of all four different subgroups (green stars, green circles, orange stars, and orange circles) of entities.

As in this setting the real quality $q$ is independent from \textit{shape} and \textit{color}, each group $g$ would be proportionally represented in an unbiased selection of the top $k\%$ candidates. That is, the chances of an entity from group $g$ to appear in the top $k\%$ would be equal to its share among all entities: $N_g / N_{total}$, where $N_g$ is the size of group $g$ and $N_{total}$ is the total number of entities.
Through the introduced bias, selection based on the perceived quality, however, $\hat{q}$ can lead to different outcomes.
Thus, we can compute a \emph{representation bias} $B_g$, i.e. the under- or over-representation of a certain group $g$ in biased selections as the relative change in chances for its members to appear in top $k\%$ positions. If the proportion of entities from group $g$ in top $k\%$ positions is higher than $N_g / N_{total}$ then the group is over-represented, if lower -- than the group is under-represented

To avoid such misrepresentation and achieve statistical parity, a common solution is to apply quota-based debiasing by allocating a proportional number of positions for each group in the top $k\%$ rankings and filling them with the candidates with highest $\hat{q}$. 

Quota-based debiasing is widely used in policy- and decision-making, and underpins algorithmic debiasing \cite{zehlike2017fa}. In practice often only a single attribute is used for debiasing and other relevant attributes are either unknown or ignored. 

We demonstrate next that this could lead to unintended consequences. 

Figure \ref{fig:image} shows example results for  $f = 0.2$, $d_{color} = -0.5$ (i.e. greens are perceived as having better quality) and $d_{shape} = 1.5$ (i.e. stars are perceived as having lower quality). If only color is considered, then debiasing appears to work as intended by successfully eliminating underrepresentation: Greens perceived quality is lower than that of the orange entities (Fig. \ref{fig:image}a). As a result, they are underrepresented in top $k\%$ if the selection is blind towards all attributes but quota-based debiasing on \textit{color} successfully fixes that (Fig. \ref{fig:image}b). 
However, the apparent bias against greens is contradicts the mechanism generating the data -- by construction a green instance is at an advantage compared to an orange instance with the same shape. The appearance of a bias against green is explained by the fact that greens disproportionately consist of stars (Fig. \ref{fig:image}c) and stars are at a larger disadvantage, i.e. the penalty for stars of $1.5$ standard deviations is three times larger than advantage for green entities.

As a consequence, quota-based debiasing would in this particular example improve the position of green circles that are already the most advantaged group and worsen the position of orange stars that are already the most disadvantaged groups (Fig. \ref{fig:image}d). In detail, this situation arises in our model when $d_{shape} > - d_{color} / (1 - 2f)$ (see Appendix). This illustrates the emergence of a \emph{Debiasing Paradox}, where debiasing can decrease the representation of already underrepresented groups.

\section*{Real-world data sets}
We use several data sets from different domains to illustrate this effect in real-world settings. 

We compute the changes in representation of various subgroups for the original distributions along with the approximations from the theoretical model. For model approximations, we fixed the attributes in the original data set and model $q$ as normally distributed with the same mean and standard deviation as in the original data set. Biases $d_{color}$ and $d_{shape}$ were computed as differences in average values of $\hat{q}$ for different groups. We find that model approximations lead to qualitatively the same results for the changes in representations of groups (decreased on increased representation) implying the relevance of the model for a wide range of real-world distributions (Fig.~\ref{fig:real_world}).

\subsection{Educational outcomes}
The dataset Russian Longitudinal Panel Study of Educational and Occupational Trajectories (TrEC) \cite{malik2019russian} ($N = 2,751$) contains information about educational outcomes of students measured by PISA scores \cite{oecd2014apisa} along with contextual information about them. 
We use PISA scores as perceived quality $\hat{q}$, being from a large city as a visible attribute (advantage) and having higher socioeconomic status as a hidden attribute (advantage). These attributes are correlated in the data set with Pearson's $r = 0.10$, $P < 10^{-6}$. 
The potential intervention could be quotas for nonresident students in the best universities that are typically located in large cities. We find that debiasing on city size makes representation for students with higher socioeconomic status less biased (stars in Fig. \ref{fig:real_world}e). However, students from smaller cities with lower socioeconomic status who were already the most underrepresented group become even more underrepresented (green circles in Fig. \ref{fig:real_world}e). 

\subsection{Wages}
Next, we use a public data set on wages of City of Chicago employees \cite{chicago2019data} (N = 5779). The hourly rate was used as a perceived quality $\hat{q}$, the visible attribute is gender (being a man is an advantage), the hidden attribute is having a full-time job (advantage). These attributes are highly correlated (Pearson's $r = 0.62,$ $P < 10^{-6}$).
A potential intervention could be quotas for women that guarantee proportional representation at the highest paying positions. We find that in this case part-time working men (orange stars in Fig. \ref{fig:real_world}f) would become even more underrepresented than before.

\subsection{Scientific citations}
Furthermore, we investigate data from ``A standardized citation metrics author database annotated for scientific field'' \cite{ioannidis2019standardized} (N = 80,210).  The total number of citations from 1996 to 2017 was used as a perceived quality $\hat{q}$, gender (being male is an advantage) as the visible attribute, and being from an US institution (advantage) as the hidden attribute. These attributes are negatively correlated in the database: Pearson's $r = -0.04$, $P < 10^{-6}$, i.e., the proportion of women is larger for US institutions than for non-US institutions. Potential intervention could be gender quotas in search results sorted by number of citations (e.g. google scholar) that ensure proportional representation of women. We find that such debiasing would decrease the representation of non-US men who were already underrepresented (orange stars in Fig. \ref{fig:real_world}g) and US women would become overrepresented (green circles in Fig. \ref{fig:real_world}g). Note that as the number of citations is not normally distributed as assumed by the theoretical model, this value was log-transformed for model approximation (Fig. \ref{fig:real_world}k). 

\subsection{Recidivism risk score}
\begin{figure}[ht!]
\centering
\includegraphics[width=\linewidth]{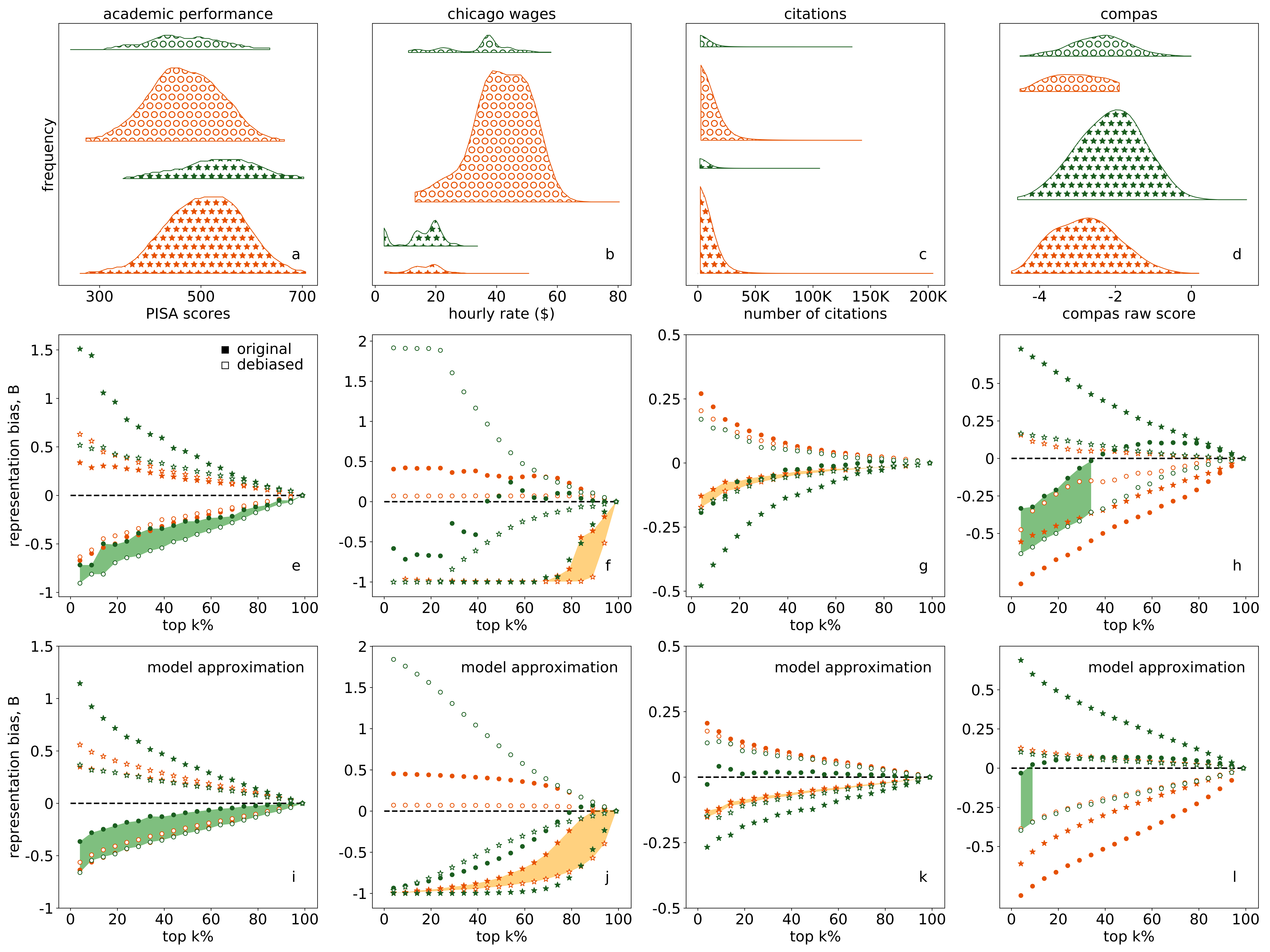}
\caption{\textbf{Illustration of the Debiasing Paradox with real-world data along with model approximations}. In all cases, debiasing decreases representation for some of the already underrepresented groups (panels e-h). In some cases debiasing decreases the representation of the most underrepresented group (panel e). Model approximation leads to qualitatively the same results for the changes in representations of groups (decreased on increased representation) implying relevance of our model for the wide range of real-world distributions. }
\label{fig:real_world}
\end{figure}

We use data from COMPAS Recidivism Risk Score data set \cite{compas2020data}. The attribute Raw score was used as a perceived quality $\hat{q}$,  being African American (higher score) as visible attribute, and being a man (higher score) as hidden attribute. These attributes are weakly, but significantly correlated in the data set, Pearson's $r = 0.05$, $P < 10^{-6}$. Potential intervention could be an adjustment of scores such that there is no bias for gender. We find that in this case non-African American man who are already underrepresented in the top positions (green circles in Fig. \ref{fig:real_world}h)  would become even less represented. Note that in this case being underrepresented is a positive outcome for a person as higher score decreases the probability of defendant to be released before trial.

\subsection*{Effects on the overall fairness of selection}
The goal of a selection process is typically to maximize the average quality $q$ of selected people. If the real quality is uncorrelated to other attributes that only influence the perception, then the unbiased selection of top $k\%$ ranking candidates would achieve this goal as it selects candidates with highest possible $q$. 
A biased selection is based on perceived quality $\hat{q}$ instead.
Thus, the average performance of selected candidates would be typically lower. 
Thus, being aware of alternative measure definitions~\cite{mehrabi2019survey,saxena2019fairness}, we quantify fairness of a selection $F_k$ as the difference in quality between (i) an average person at top $k\%$ positions of a biased selection and (ii) an average person at top $k\%$ positions of an unbiased selection. As unbiased selection maximizes the quality of selected participants, the maximum value of $F_k$ is zero.

The Debiasing Paradox raises the question whether the overall fairness of a selection is improved after debiasing. To answer this question, we explore the changes in average quality of selected candidates for different values of $d_{color}$ and $d_{shape}$. 
While the real quality in real world data is almost always not observable, we can study this effect with the introduced theoretic model.

\begin{figure}[ht!]
\centering
\includegraphics[width=\linewidth]{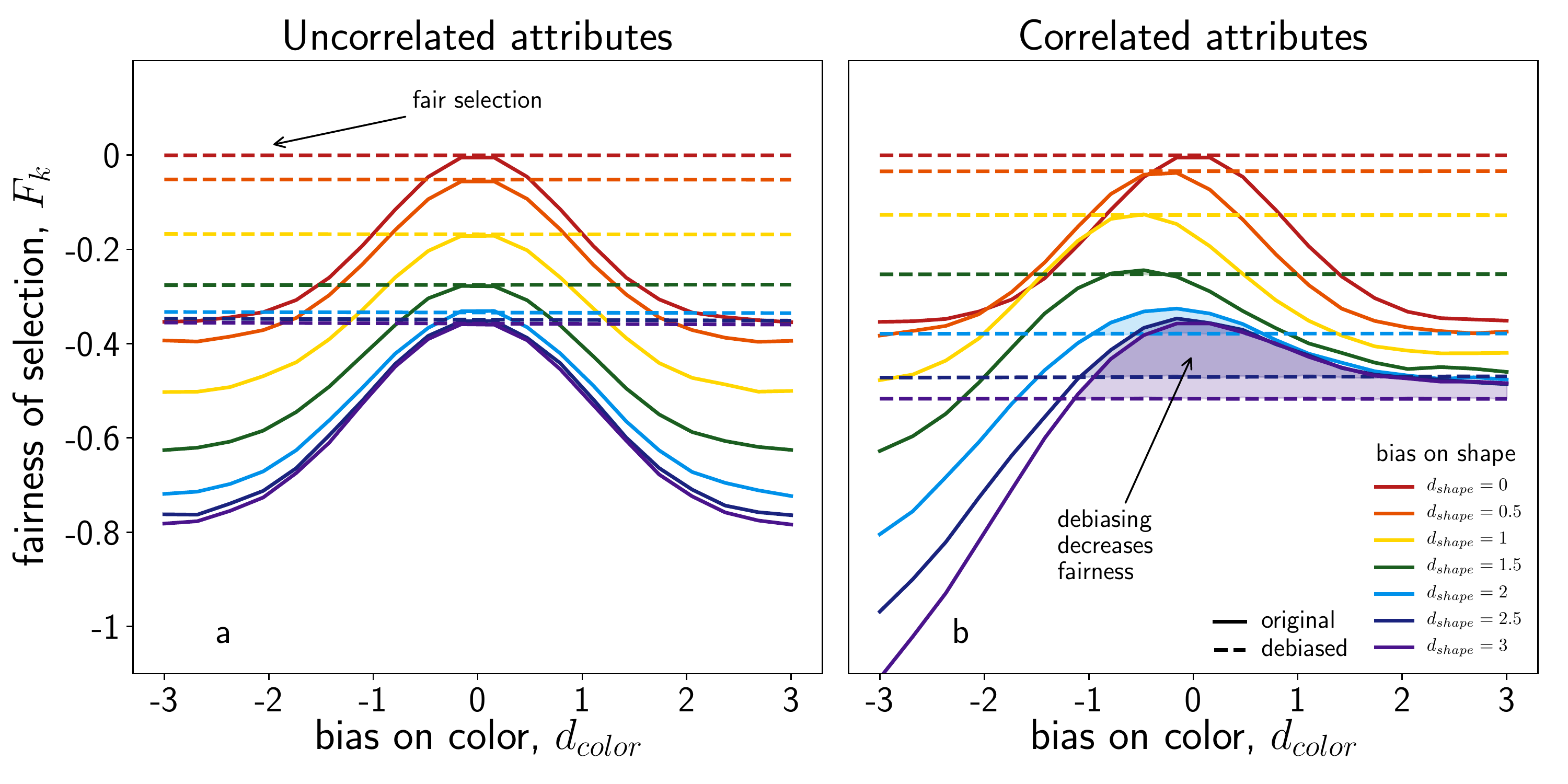}
\caption{\textbf{The effects of debiasing on the quality of selected candidates for uncorrelated (a) and correlated (b) attributes.} When shape and color are uncorrelated and there is no bias for shape (red solid line on panel a) then debiasing on color successfully maximizes the quality of selected candidates. If there exists bias for shape then the maximum value is not achieved but quality is still improved. However, if two attributes are correlated then in some cases overall quality \emph{decreases} (shaded regions on panel b).}
\label{fig:overall}
\end{figure}

Figure \ref{fig:overall}a demonstrates changes in the quality of selected candidates if color and shape are uncorrelated ($f = 0.5$, $k = 0.2$). If there is no bias for shape (red solid line) then debiasing on color successfully maximizes the real quality of candidates (red dashed line). In other cases, debiasing would not achieve the maximum value, but would still improve the quality of selected candidates. 

Figure \ref{fig:overall}b shows results for the case that color and shape are correlated ($f = 0.2$, $k = 0.2$). It can be observed that there are cases, in which the quality of participants \emph{decreases after debiasing}. One scenario where this would happen is when there is no bias on color but large enough bias on shape. This would for instance happen when debiasing is based on gender while gendered behaviour and not gender itself is penalized. In this case introducing quotas would make the selection less fair if the penalty is large enough.

\section*{Discussion}
Quota-based debiasing is an effective way to remove bias with respect to a single binary attribute.
Our work demonstrates the potential negative side-effects of quotas on subgroups. These effects can appear in situations with incomplete knowledge, i.e. when some relevant attributes are unknown or ignored.
Well-known examples of such situations include gender disparities in UC Berkeley admissions ~\cite{bickel1975sex} or racial disparities in death penalty decisions~\cite{radelet1981racial}. 
Our work takes an action-oriented perspective towards these problems, by highlighting the potential unintended consequences of interventions such as quotas.
In particular, we show quota-based debiasing could worsen the representation of already underrepresented groups and decrease overall fairness of rankings. Our work studies these effects from a statistical point of view, but does not consider potential indirect effects and long-term consequences~\cite{bowen1998shape,francis2012using}.

While debiasing -- and specifically quota-based debiasing -- has been applied for decades, the recent rise of artificial intelligence systems amplifies and compounds this problem. Artificial intelligence systems have been shown to reproduce or even enhance biases from training data~\cite{obermeyer2019dissecting,mehrabi2019survey}. 
This calls for generally applicable solutions to avoid such biases.
Our work raises awareness that automatically applied, simplistic quotas to achieve statistical parity of groups might not be a suitable solutions, but can in some scenarios even worsen disparate representations of subgroups. This shall encourage further research on alternative domain-specific approaches, see for example ~\cite{krook2014beyond}.

Overall, this work demonstrates that quota-based debiasing warrants caution and control for various additional attributes. In some cases, it could be impossible to fix bias for one attribute without introducing bias for another. Instead of solely relying on post-hoc fixing of biases via quotas, efforts should be directed towards eliminating the root causes of biases. 

\section*{Data and code availability}
Code to reproduce the main results of the paper is available at \url{https://github.com/ibsmirnov/debiasing}

\section*{Acknowledgement}
Ivan Smirnov acknowledges support from the Basic Research Program of the National Research University Higher School of Economics.

\bibliographystyle{unsrt}
\bibliography{references}

\clearpage
\beginsupplement
\section*{Appendix}

\subsection*{Changes in representation of the most underrepresented group}
In this section, we explore how the representation of the most underrepresented group changes depending on the parameter values. 
Table \ref{tab:summary} summarizes biases for and sizes of all four subgroups.

\begin{table}[ht]
\caption{Group summary}
\begin{tabular}{l|c|c|c|c}
& \text{\Large\color{orange}$\star$} & \text{\Large\color{ForestGreen}$\star$} & \text{\Large\color{orange}$\bullet$} & \text{\Large\color{ForestGreen}$\bullet$} \\
\hline
group size & $fN$ & $(1-f)N$ & $(1-f)N$ & $fN$ \\
\hline
bias & $-d_{shape}$ & $-d_{shape} - d_{color}$ & 0  & $-d_{color}$  
\end{tabular}
\label{tab:summary}
\end{table}

\begin{figure}[ht]
\centering
\includegraphics[width=0.5\linewidth]{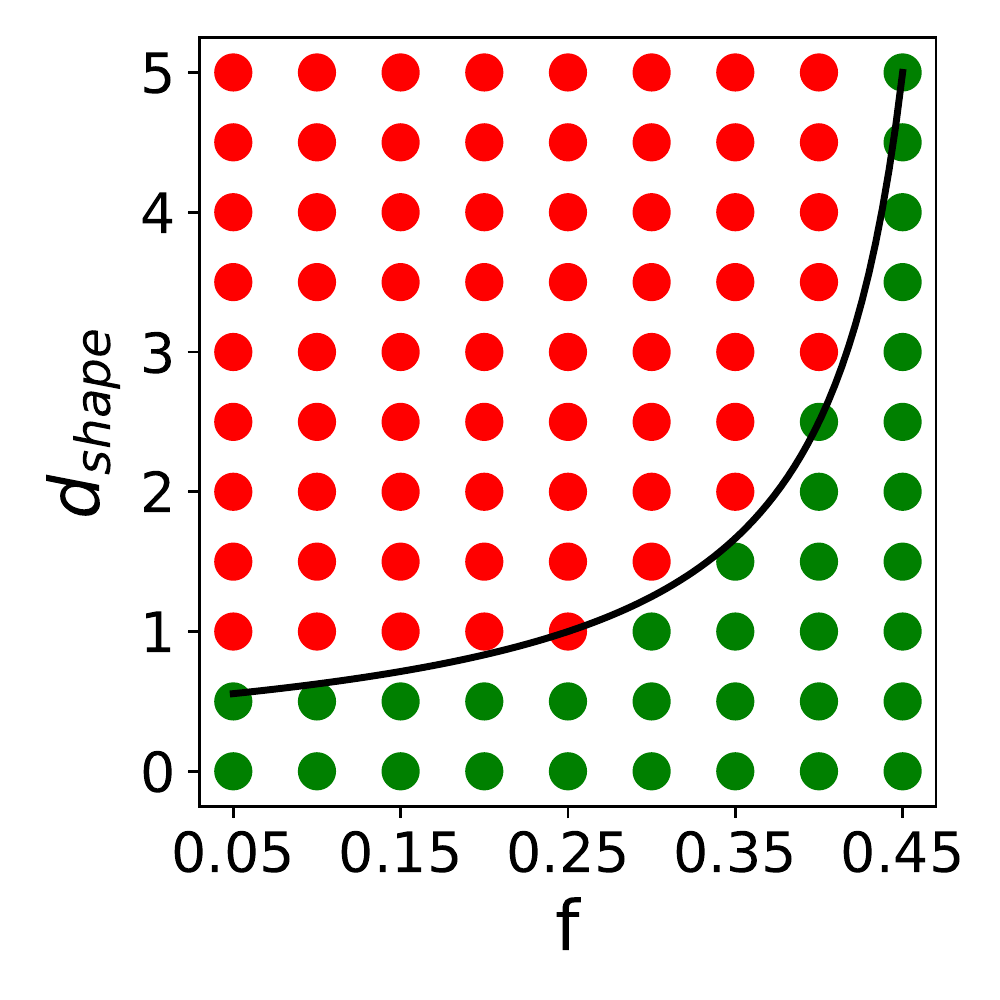}
\caption{\textbf{Changes in representation of the most underrepresented group}. Green color corresponds to the increased representation and red to the decreased representation. The black line corresponds to the analytically derived curve $d_{shape} = -d_{color} / (1 - 2f)$}
\label{fig:simulation}
\end{figure}

Let's assume that $d_{shape} > 0$ and $d_{color} < 0$, i.e. there is penalty for being star and advantage for being green (or penalty for being orange). While there is a penalty for being orange, on aggregated level oranges might seem to have an advantage because there are many orange circles and circles have an advantage. The average bias for orange entities would be $d_{orange} = (-d_{shape} \times fN + 0 \times (1 - f)N)) / N = - d_{shape} \times f$. The average bias for green entities would be $d_{green} = ((-d_{shape} - d_{color}) \times (1-f)N - d_{color} \times fN) / N = -d_{shape} - d_{color} + d_{shape} \times f$. On aggregated level oranges would have an advantage when $d_{orange} > d_{green}$, or $-d_{shape} \times f > -d_{shape} - d_{color} + d_{shape} \times f$, or $d_{shape} > -d_{color} / (1 - 2f)$. In this case, the debiasing would improve the representation of green entities and worsen representation of orange entities, including the orange stars -- already the most underrepresented group. Figure \ref{fig:simulation} shows the results of simulation for various pairs of $d_{shape}$ and $f$ with fixed $d_{color} = -0.5$ ($10,000$ simulations for each pair). The black line corresponds to the $d_{shape} = -d_{color} / (1 - 2f)$ curve.

\subsection*{Data}
\subsubsection*{Educational outcomes}
We use data from a representative Russian panel study titled "Trajectories in Education and Careers" (TrEC) \cite{malik2019russian}. The study tracks 4,400 students from 42 Russian regions who took the PISA test in 2012 \cite{oecd2014apisa}. We used PISA reading scores as a measure of students’ academic performance. PISA defines reading literacy as "understanding, using, reflecting on and engaging with written texts in order to achieve one’s goals, to develop one’s knowledge and potential, and to participate in society" and considers it a foundation for achievement in other subject areas within the educational system and also a prerequisite for successful participation in most areas of adult life \cite{oecd2014apisa}. PISA scores are scaled so that the OECD average is 500 and the standard deviation is 100, while every 40 score points roughly correspond to the equivalent of one year of formal schooling \cite{oecd2014apisa}. 

The data set contains information about socioeconomic status of students measured by the PISA index of economic, social and cultural status (ESCS). ESCS is derived from the the three indices measuring highest occupational status of parents, highest education level of parents in years of education, and home possessions (each variable has an OECD mean of zero and a standard deviation of one) \cite{oecd2014apisa}. We binarize this variable by defining the top half of students from the data set as having higher socioeconomic status and the bottom half of students as having lower socioeconomic status.

For each participant, the type of settlement where they live is known. With the largest category including cities with population of more than $680,000$. We consider this category as large cities ($20\%$ of participants live in such large cities).

After removing observations with missing values the final size of sample is $2,751$.

\subsubsection*{Wages}
We use public data set on wages of City of Chicago employees \cite{chicago2019data}. We use a binary 'part or full time' variable from the data set along with 'hourly rate' variable as hidden attribute and perceived quality $q$, respectively. We infer potential gender of a person from their first names using \emph{gender\_guesser} python package.
We acknowledge that this procedure comes with a risk of inaccuracy, specifically for individuals with a non-western cultural background.~\cite{karimi2016inferring}.

\subsubsection*{Scientific citations}
We employ data from ``A standardized citation metrics author database annotated for scientific field'' \cite{ioannidis2019standardized}. We use a 'nc9617 (ns)' variable (total number of citations from 1996 to 2017 excluding self-citations) as perceived quality. The variable 'cntry' (country associated with most recent institution) was binarized into being from USA institution and used as hidden attribute. The likely gender was inferred based on the \emph{gender\_guesser} python package and was used as the visible attribute. 

\subsubsection*{Recidivism risk score}
We studied data from COMPAS Recidivism Risk Score data set \cite{compas2020data}. Correctional Offender Management Profiling for Alternative Sanctions (COMPAS) is a widely used criminal risk assessment tool \cite{dressel2018accuracy} in the USA. The software predicts a defendant’s risk of committing a misdemeanor or felony within 2 years of assessment from 137 features about an individual and the individual’s past criminal record \cite{dressel2018accuracy}. Through a public records request, ProPublica obtained two years worth of COMPAS scores from the Broward County Sheriff’s Office in Florida and made it publicly available \cite{compas2020data}. We choose COMPAS raw score for the risk of violence as perceived quality. Not that unlike in other examples, having higher score can lead to negative consequences for a person, i.e. being overrepresented at top positions is worse for a group of people than being underrepresented. We binarize 'Ethnic Code' variable by assigning to one group people with codes 'African-American' and 'African-Am', and to another group people with codes 'Asian' and 'Caucasian'. This variable was used as a visible attribute and 'Sex Code' variable was used as a hidden attribute.

\end{document}